\pdfoutput=1
\documentclass[10pt,letter]{article}
\usepackage{dblfloatfix}

\usepackage{ifthen} % for conditional statements
\newboolean{pdflatex}
\setboolean{pdflatex}{true} % False for eps figures 

\newboolean{articletitles}
\setboolean{articletitles}{true} % False removes titles in references

\newboolean{uprightparticles}
\setboolean{uprightparticles}{false} %True for upright particle symbols

\newboolean{inbibliography}
\setboolean{inbibliography}{false} %True once you enter the bibliography

\usepackage{microtype}
\usepackage{lineno}  % for line numbering during review
\usepackage{xspace} % To avoid problems with missing or double spaces after
\usepackage{graphicx}  % to include figures (can also use other packages)
\usepackage{color}
\usepackage{colortbl}
\graphicspath{{./figs/}} % Make Latex search fig subdir for figures

\usepackage{amsmath} % Adds a large collection of math symbols
\usepackage{amssymb}
\usepackage{amsfonts}
\usepackage{upgreek} % Adds in support for greek letters in roman typeset

\newcommand*\patchAmsMathEnvironmentForLineno[1]{%
\expandafter\let\csname old#1\expandafter\endcsname\csname #1\endcsname
\expandafter\let\csname oldend#1\expandafter\endcsname\csname
end#1\endcsname
 \renewenvironment{#1}%
   {\linenomath\csname old#1\endcsname}%
   {\csname oldend#1\endcsname\endlinenomath}%
}
\newcommand*\patchBothAmsMathEnvironmentsForLineno[1]{%
  \patchAmsMathEnvironmentForLineno{#1}%
  \patchAmsMathEnvironmentForLineno{#1*}%
}
\AtBeginDocument{%
\patchBothAmsMathEnvironmentsForLineno{equation}%
\patchBothAmsMathEnvironmentsForLineno{align}%
\patchBothAmsMathEnvironmentsForLineno{flalign}%
\patchBothAmsMathEnvironmentsForLineno{alignat}%
\patchBothAmsMathEnvironmentsForLineno{gather}%
\patchBothAmsMathEnvironmentsForLineno{multline}%
}

\usepackage{hyperref}    % Hyperlinks in references
\usepackage[all]{hypcap} % Internal hyperlinks to floats.

\def\lhcb {\mbox{LHCb}\xspace}

\ifthenelse{\boolean{uprightparticles}}%
{

 \def\Ppi         {\ensuremath{\uppi}\xspace}

 \def\Ppsi        {\ensuremath{\uppsi}\xspace}

 \def\PDelta      {\ensuremath{\Delta}\xspace}                 
 \def\PXi      {\ensuremath{\Xi}\xspace}                 
 \def\PLambda      {\ensuremath{\Lambda}\xspace}                 
 \def\PSigma      {\ensuremath{\Sigma}\xspace}                 
 \def\POmega      {\ensuremath{\Omega}\xspace}                 
 \def\PUpsilon      {\ensuremath{\Upsilon}\xspace}                 
 
 \def\PB      {\ensuremath{\mathrm{B}}\xspace}                 
                  
 \def\PD      {\ensuremath{\mathrm{D}}\xspace}

 \def\PJ      {\ensuremath{\mathrm{J}}\xspace}                 
 \def\PK      {\ensuremath{\mathrm{K}}\xspace}

 \def\Pb      {\ensuremath{\mathrm{b}}\xspace}                 
 \def\Pc      {\ensuremath{\mathrm{c}}\xspace}

 \def\Pi      {\ensuremath{\mathrm{i}}\xspace}

 \def\Ps      {\ensuremath{\mathrm{s}}\xspace}

}
{

 \def\Ppi         {\ensuremath{\pi}\xspace}

 \def\Ppsi        {\ensuremath{\psi}\xspace}                 
                  
 \mathchardef\PDelta="7101
 \mathchardef\PXi="7104
 \mathchardef\PLambda="7103
 \mathchardef\PSigma="7106
 \mathchardef\POmega="710A
 \mathchardef\PUpsilon="7107
                  
 \def\PB      {\ensuremath{B}\xspace}                 
                  
 \def\PD      {\ensuremath{D}\xspace}

 \def\PJ      {\ensuremath{J}\xspace}                 
 \def\PK      {\ensuremath{K}\xspace}

 \def\Pb      {\ensuremath{b}\xspace}                 
 \def\Pc      {\ensuremath{c}\xspace}

 \def\Pi      {\ensuremath{i}\xspace}

 \def\Ps      {\ensuremath{s}\xspace}

}

\def\squark    {\ensuremath{\Ps}\xspace}

\def\cquark    {\ensuremath{\Pc}\xspace}

\def\bquark    {\ensuremath{\Pb}\xspace}

\def\pion  {\ensuremath{\Ppi}\xspace}
\def\piz   {\ensuremath{\pion^0}\xspace}

\def\pip   {\ensuremath{\pion^+}\xspace}
\def\pim   {\ensuremath{\pion^-}\xspace}

\def\kaon  {\ensuremath{\PK}\xspace}
  \def\Kbar  {\kern 0.2em\overline{\kern -0.2em \PK}{}\xspace}

\def\Kp    {\ensuremath{\kaon^+}\xspace}
\def\Km    {\ensuremath{\kaon^-}\xspace}

  \def\Dbar    {\kern 0.2em\overline{\kern -0.2em \PD}{}\xspace}
\def\D       {\ensuremath{\PD}\xspace}

\def\Dz      {\ensuremath{\D^0}\xspace}
\def\Dzb     {\ensuremath{\Dbar^0}\xspace}

\def\Dstarp  {\ensuremath{\D^{*+}}\xspace}

\def\Ds      {\ensuremath{\D^+_\squark}\xspace}

\def\B       {\ensuremath{\PB}\xspace}
\def\Bbar    {\ensuremath{\kern 0.18em\overline{\kern -0.18em \PB}{}}\xspace}

\def\Bz      {\ensuremath{\B^0}\xspace}

\def\Bs      {\ensuremath{\B^0_\squark}\xspace}

\def\jpsi     {\ensuremath{{\PJ\mskip -3mu/\mskip -2mu\Ppsi\mskip 2mu}}\xspace}

  \def\Y#1S{\ensuremath{\PUpsilon{(#1S)}}\xspace}% no space before {...}!

\def\Lbar {\ensuremath{\kern 0.1em\overline{\kern -0.1em\PLambda}}\xspace}

\newcommand{\decay}[2]{\ensuremath{#1\!\to #2}\xspace}         % {\Pa}{\Pb \Pc}

\def\to                 {\ensuremath{\rightarrow}\xspace}

\def\CP                {\ensuremath{C\!P}\xspace}

\def\AT#1     {\ensuremath{A_{\mathrm{T}}^{#1}}\xspace}           % 2

\def\C#1      {\ensuremath{\mathcal{C}_{#1}}\xspace}                       % 9
\def\Cp#1     {\ensuremath{\mathcal{C}_{#1}^{'}}\xspace}                    % 7
\def\Ceff#1   {\ensuremath{\mathcal{C}_{#1}^{\mathrm{(eff)}}}\xspace}        % 9  
\def\Cpeff#1  {\ensuremath{\mathcal{C}_{#1}^{'\mathrm{(eff)}}}\xspace}       % 7
\def\Ope#1    {\ensuremath{\mathcal{O}_{#1}}\xspace}                       % 2
\def\Opep#1   {\ensuremath{\mathcal{O}_{#1}^{'}}\xspace}                    % 7

\def\ycp        {\ensuremath{y_{\CP}}\xspace}
\def\agamma     {\ensuremath{A_{\Gamma}}\xspace}
\newcommand{\tev}{\ifthenelse{\boolean{inbibliography}}{\ensuremath{~T\kern -0.05em eV}\xspace}{\ensuremath{\mathrm{\,Te\kern -0.1em V}}\xspace}}
\newcommand{\gev}{\ensuremath{\mathrm{\,Ge\kern -0.1em V}}\xspace}
\newcommand{\mev}{\ensuremath{\mathrm{\,Me\kern -0.1em V}}\xspace}
\newcommand{\kev}{\ensuremath{\mathrm{\,ke\kern -0.1em V}}\xspace}
\newcommand{\ev}{\ensuremath{\mathrm{\,e\kern -0.1em V}}\xspace}
\newcommand{\gevc}{\ensuremath{{\mathrm{\,Ge\kern -0.1em V\!/}c}}\xspace}
\newcommand{\mevc}{\ensuremath{{\mathrm{\,Me\kern -0.1em V\!/}c}}\xspace}
\newcommand{\gevcc}{\ensuremath{{\mathrm{\,Ge\kern -0.1em V\!/}c^2}}\xspace}
\newcommand{\gevgevcccc}{\ensuremath{{\mathrm{\,Ge\kern -0.1em V^2\!/}c^4}}\xspace}
\newcommand{\mevcc}{\ensuremath{{\mathrm{\,Me\kern -0.1em V\!/}c^2}}\xspace}

\def\invfb   {\ensuremath{\mbox{\,fb}^{-1}}\xspace}

\def\ps   {\ensuremath{{\rm \,ps}}\xspace}
\def\fs   {\ensuremath{\rm \,fs}\xspace}

\newcommand{\chisq}{\ensuremath{\chi^2}\xspace}

\newcommand{\chisqip}{\ensuremath{\chi^2_{\rm IP}}\xspace}

\def\gsim{{~\raise.15em\hbox{$>$}\kern-.85em
          \lower.35em\hbox{$\sim$}~}\xspace}
\def\lsim{{~\raise.15em\hbox{$<$}\kern-.85em
          \lower.35em\hbox{$\sim$}~}\xspace}

\def\pt         {\mbox{$p_{\rm T}$}\xspace}

\def\tell1  {TELL1\xspace}
\def\ukl1   {UKL1\xspace}

\usepackage{cite} % Allows for ranges in citations
\usepackage{mciteplus}

\def\doc {\mbox{Letter}\xspace}

\newcommand{\deltam}{\ensuremath{\Delta m}\xspace}

\usepackage{longtable} % only for template; not usually to be used in PAPERs

\twocolumn

\begin{document}

%%%%%%%%%%%%%%%%%%%%%%%%%
%%%%% Title     %%%%%%%%%
%%%%%%%%%%%%%%%%%%%%%%%%%
\renewcommand{\thefootnote}{\fnsymbol{footnote}}
\setcounter{footnote}{1}

% %%%%%%% CHOOSE TITLE PAGE--------
\onecolumn
% \input{title-LHCb-ANA}
%\input{title-LHCb-CONF}
% $Id: title-LHCb-PAPER.tex 42099 2013-10-03 08:17:11Z gersabec $
% ===============================================================================
% Purpose: LHCb-PAPER journal paper title page template
% Author: 
% Created on: 2010-09-25
% ===============================================================================

%%%%%%%%%%%%%%%%%%%%%%%%%
%%%%%  TITLE PAGE  %%%%%%
%%%%%%%%%%%%%%%%%%%%%%%%%
\begin{titlepage}
\pagenumbering{roman}

% Header ---------------------------------------------------
\vspace*{-1.5cm}
\centerline{\large EUROPEAN ORGANIZATION FOR NUCLEAR RESEARCH (CERN)}
\vspace*{1.5cm}
\hspace*{-0.5cm}
\begin{tabular*}{\linewidth}{lc@{\extracolsep{\fill}}r}
\ifthenelse{\boolean{pdflatex}}% Logo format choice
{\vspace*{-2.7cm}\mbox{\!\!\!\includegraphics[width=.14\textwidth]{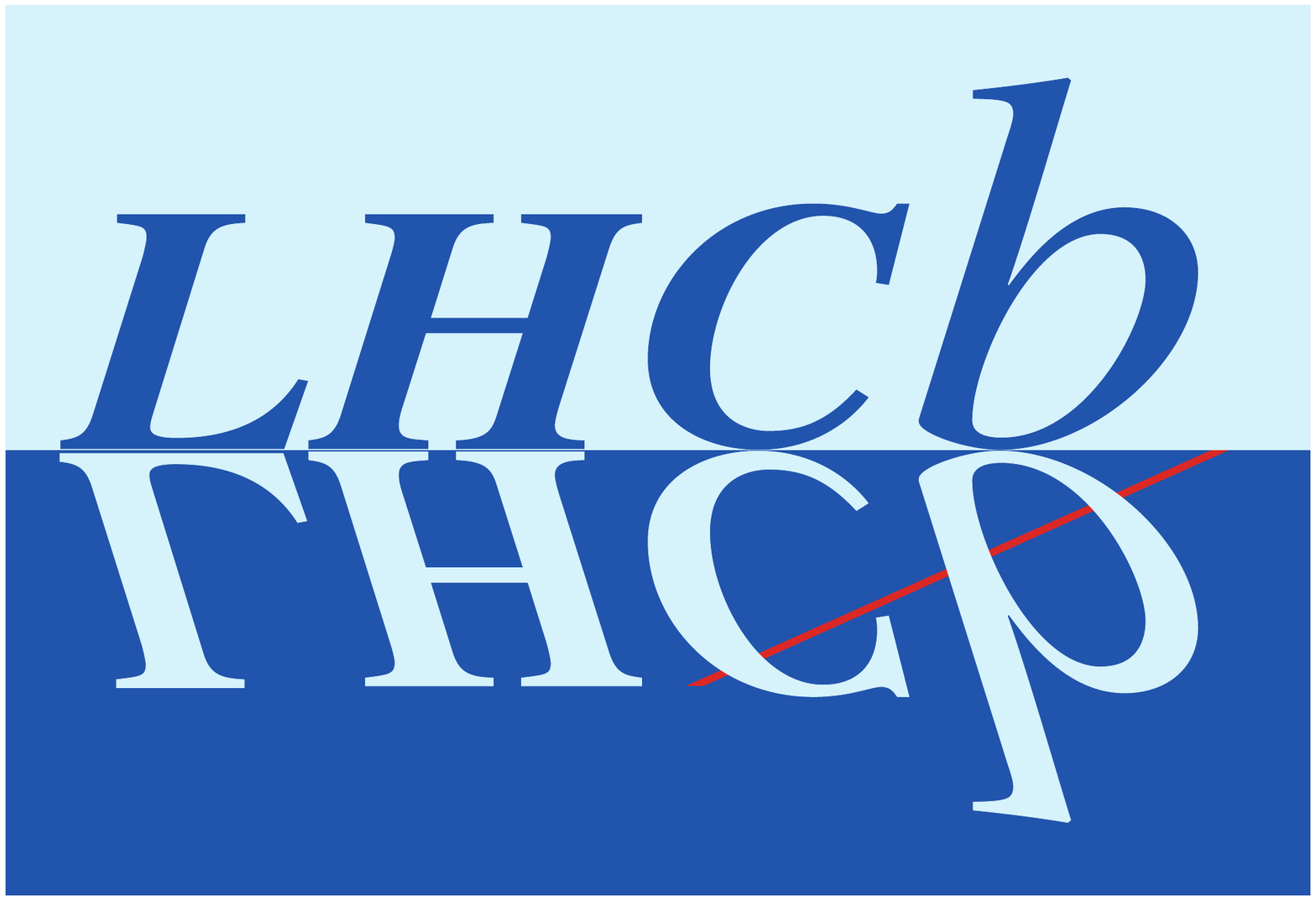}} & &}%
{\vspace*{-1.2cm}\mbox{\!\!\!\includegraphics[width=.12\textwidth]{lhcb-logo.eps}} & &}%
\\
 & & CERN-PH-EP-2013-180 \\  % ID 
 & & LHCb-PAPER-2013-054 \\  % ID 
 & & \today \\ % Date - Can also hardwire e.g.: 23 March 2010
% not in paper \hline
\end{tabular*}

\vspace*{4.0cm}

% Title --------------------------------------------------
{\bf\boldmath\huge
\begin{center}
  Measurements of indirect \CP asymmetries in \decay{\Dz}{\Km\Kp} and \decay{\Dz}{\pim\pip} decays
\end{center}
}

\vspace*{2.0cm}

% Authors -------------------------------------------------
\begin{center}
The LHCb collaboration\footnote{Authors are listed on the following pages.}
\end{center}

\vspace{\fill}

% Abstract -----------------------------------------------
\begin{abstract}
  \noindent
  A study of indirect \CP violation in \Dz mesons through the determination of the parameter \agamma is presented using a data sample of $pp$ collisions, corresponding to an integrated luminosity of $1.0\invfb$, collected with the \lhcb detector and recorded at the centre-of-mass energy of $7\tev$ at the LHC.
  The parameter \agamma is the asymmetry of the effective lifetimes measured in decays of \Dz and \Dzb mesons to the \CP eigenstates $\Km\Kp$ and $\pim\pip$.
  Fits to the data sample yield $\agamma(KK)=(-0.35\pm0.62\pm0.12)\times 10^{-3}$ and $\agamma(\pi\pi)=(0.33\pm1.06\pm0.14)\times 10^{-3}$, where the first uncertainties are statistical and the second systematic.
  The results represent the world's best measurements of these quantities.
  They show no difference in \agamma between the two final states and no indication of \CP violation.
\end{abstract}

\vspace*{2.0cm}

\begin{center}
  Accepted for publication in Phys.~Rev.~Lett.
\end{center}

\vspace{\fill}

{\footnotesize 
\centerline{\copyright~CERN on behalf of the \lhcb collaboration, license \href{http://creativecommons.org/licenses/by/3.0/}{CC-BY-3.0}.}}
\vspace*{2mm}

\end{titlepage}

%%%%%%%%%%%%%%%%%%%%%%%%%%%%%%%%
%%%%%  EOD OF TITLE PAGE  %%%%%%
%%%%%%%%%%%%%%%%%%%%%%%%%%%%%%%%

%  empty page follows the title page ----
\newpage
\setcounter{page}{2}
\mbox{~}
\newpage

% Author List ----------------------------
%  You need to get a new author list!
%%%%%%%%%%%%%%%%%%%%%%%%%%%%%%%%%%%%%%%%%%
\centerline{\large\bf LHCb collaboration}
\begin{flushleft}
\small
R.~Aaij$^{40}$, 
B.~Adeva$^{36}$, 
M.~Adinolfi$^{45}$, 
C.~Adrover$^{6}$, 
A.~Affolder$^{51}$, 
Z.~Ajaltouni$^{5}$, 
J.~Albrecht$^{9}$, 
F.~Alessio$^{37}$, 
M.~Alexander$^{50}$, 
S.~Ali$^{40}$
% ...

{\footnotesize \it
% ...
$ ^{5}$Clermont Universit\'{e}, Universit\'{e} Blaise Pascal, CNRS/IN2P3, LPC, Clermont-Ferrand, France\\
$ ^{6}$CPPM, Aix-Marseille Universit\'{e}, CNRS/IN2P3, Marseille, France\\
% ...
$ ^{9}$Fakult\"{a}t Physik, Technische Universit\"{a}t Dortmund, Dortmund, Germany\\
% ...
$ ^{36}$Universidad de Santiago de Compostela, Santiago de Compostela, Spain\\
$ ^{37}$European Organization for Nuclear Research (CERN), Geneva, Switzerland\\
% ...
$ ^{40}$Nikhef National Institute for Subatomic Physics, Amsterdam, The Netherlands\\
% ...
$ ^{45}$H.H. Wills Physics Laboratory, University of Bristol, Bristol, United Kingdom\\
% ...
$ ^{50}$School of Physics and Astronomy, University of Glasgow, Glasgow, United Kingdom\\
$ ^{51}$Oliver Lodge Laboratory, University of Liverpool, Liverpool, United Kingdom\\
% ...
}
\end{flushleft}
%%%%%%%%%%%%%%%%%%%%%%%%%%%%%%%%%%%%%%%%%%

\cleardoublepage

\twocolumn
% %%%%%%%%%%%%% ---------

\renewcommand{\thefootnote}{\arabic{footnote}}
\setcounter{footnote}{0}

%%%%%%%%%%%%%%%%%%%%%%%%%%%%%%%%
%%%%%  Table of Content   %%%%%%
%%%%%%%%%%%%%%%%%%%%%%%%%%%%%%%%
%%%% Uncomment next 2 lines if desired
%\tableofcontents
%\cleardoublepage

%%%%%%%%%%%%%%%%%%%%%%%%%
%%%%% Main text %%%%%%%%%
%%%%%%%%%%%%%%%%%%%%%%%%%

\pagestyle{plain} % restore page numbers for the main text
\setcounter{page}{1}
\pagenumbering{arabic}

%% Uncomment during review phase. 
%% Comment before a final submission.
% \linenumbers

% You can include short sections directly in the main tex file.
% However, for larger papers it is desirable to split the text into
% several semiautonomous files, which can be revised independently.
% This is especially useful when developing a document in
% collaboration with several people, since then different parts can be
% edited independently.  This type of file organization is shown here.
% 

% $Id: body.tex 42185 2013-10-06 16:45:36Z gersabec $

The asymmetry under simultaneous charge and parity transformation (\CP violation) has driven the understanding of electroweak interactions since its discovery in the kaon system~\cite{Christenson:1964fg}.
\CP violation was subsequently discovered in the \Bz and \Bs systems~\cite{Aubert:2001nu,Abe:2001xe,LHCb-PAPER-2013-018}.
Charmed mesons form the only neutral meson-antimeson system in which \CP violation has yet to be observed unambiguously.
This system is the only one in which mesons of up-type quarks participate in matter-antimatter transitions, a loop-level process in the Standard Model (SM).
This charm mixing process has recently been observed for the first time unambiguously in single measurements~\cite{LHCb-PAPER-2012-038,Aaltonen:2013pja,LHCb-PAPER-2013-053}.
The theoretical calculation of charm mixing and \CP violation is challenging for the charm quark~\cite{Bobrowski:2010xg,Georgi:1992as,Ohl:1992sr,Bigi:2000wn,Lenz:2013aua}.
Significant enhancement of mixing or \CP violation would be an indication of physics beyond the SM.

The mass eigenstates of the neutral charm meson system, $|\PD_{1,2}\rangle$, with masses $m_{1,2}$ and decay widths $\Gamma_{1,2}$, can be expressed as linear combinations of the flavour eigenstates, $|\Dz\rangle$ and $|\Dzb\rangle$, as $|\PD_{1,2}\rangle=p|\Dz\rangle\pm{}q|\Dzb\rangle$ with complex coefficients satisfying $|p|^2+|q|^2=1$.
This allows the definition of the mixing parameters $x\equiv 2(m_2-m_1)/(\Gamma_1+\Gamma_2)$ and $y\equiv(\Gamma_2-\Gamma_1)/(\Gamma_1+\Gamma_2)$.

Non-conservation of \CP symmetry enters as a deviation from unity of $\lambda_f$, defined as
\begin{equation}
\label{eqn:charm_lambda}
\lambda_f\equiv\frac{q\bar{A}_f}{pA_f}=-\eta_{\CP}\left|\frac{q}{p}\right|\left|\frac{\bar{A}_f}{A_f}\right|e^{i\phi},
\end{equation}
where $A_f$ ($\bar{A}_f$) is the amplitude for a \Dz (\Dzb) meson decaying into a \CP eigenstate $f$ with eigenvalue $\eta_{\CP}$, and $\phi$ is the \CP-violating relative phase between $q/p$ and $\bar{A}_f/A_f$.
Direct \CP violation occurs when the asymmetry $A_d\equiv(|A_f|^2-|\bar{A}_f|^2)/(|A_f|^2+|\bar{A}_f|^2)$ is different from zero.
Indirect \CP violation comprises non-zero \CP asymmetry in mixing, $A_m\equiv(|q/p|^2-|p/q|^2)/(|q/p|^2+|p/q|^2)$ and \CP violation through a non-zero phase $\phi$.
The phase convention of $\phi$ is chosen such that, in the limit of no \CP violation, $\CP|\Dz\rangle=-|\Dzb\rangle$.
In this convention \CP conservation leads to $\phi=0$ and $|\PD_1\rangle$ being \CP-odd.

%%%%%%%%%%%%%%%%%%%%%%%%%%%%%%%%%%%%%%%%%%%%%%%%%%%%%%%%%%%%%%%%%%%%%%%%%%%%%%%%%%%%%%%%%

The asymmetry of the inverse of effective lifetimes in decays of \Dz (\Dzb) mesons into \CP-even final states, $\hat{\Gamma}$ ($\hat{\bar{\Gamma}}$), leads to the observable \agamma defined as
\begin{equation}
\agamma\equiv\frac{\hat{\Gamma} - \hat{\bar{\Gamma}}}{\hat{\Gamma} + \hat{\bar{\Gamma}}} \approx \eta_{\CP}\left(\frac{A_m+A_d}{2}y\cos\phi-x\sin\phi\right).
\end{equation}
This makes \agamma a measurement of indirect \CP violation, as the contributions from direct \CP violation are measured to be small~\cite{HFAG} compared to the precision on \agamma available so far~\cite{Gersabeck:2011xj}.
Here, effective lifetimes refer to lifetimes measured using a single-exponential model in a specific decay mode.
Currently available measurements of \agamma~\cite{LHCb-PAPER-2011-032,Lees:2012qh} are in agreement with no \CP violation at the per mille level~\cite{HFAG}.

This \doc reports measurements of \agamma in the \CP-even final states $\Km\Kp$ and $\pim\pip$ using $1.0\invfb$ of $pp$ collisions at $7\tev$ centre-of-mass energy at the LHC recorded with the \lhcb detector in 2011.
In the SM, the phase $\phi$ is final-state independent and thus measurements in the two final states are expected to yield the same results.
At the level of precision of the measurements presented here, differences due to direct \CP violation are negligible.
However, contributions to $\phi$ from physics beyond the SM may lead to different results.
Even small final-state differences in the phase, $\Delta\phi$, can lead to measurable effects in the observables of the order of $x\Delta\phi$, for sufficiently small phases $\phi$ in both final states~\cite{Kagan:2009gb}.
In addition, the measurements of \agamma in both final states are important to quantify the contribution of indirect \CP violation to the observable $\Delta A_{\CP}$, which measures the difference in decay-time integrated \CP asymmetry of \decay{\Dz}{\Km\Kp} to $\pim\pip$ decays~\cite{LHCb-PAPER-2011-023,LHCb-PAPER-2013-003}.

%%%%%%%%%%%%%%%%%%%%%%%%%%%%%%%%%%%%%%%%%%%%%%%%%%%%%%%%%%%%%%%%%%%%%%%%%%%%%%%%%%%%%%%%%

The \lhcb detector~\cite{Alves:2008zz} is a single-arm forward spectrometer covering the \mbox{pseudorapidity} range $2<\eta <5$, designed for the study of particles containing \bquark or \cquark quarks. 
The spectrometer dipole magnet is operated in either one of two polarities, the magnetic field vector points either up or down.
The trigger~\cite{LHCb-DP-2012-004} consists of a hardware stage, based on information from the calorimeter and muon systems, followed by a software stage, which performs a full event reconstruction.
The software trigger applies two sequential selections.
The first selection requires at least one track to have momentum transverse to the beamline, $\pt$, greater than $1.7\gevc$ and an impact parameter \chisq, $\chisqip$, greater than $16$.
The \chisqip is defined as the difference in \chisq of a given primary interaction vertex reconstructed with and without the considered track. 
This \chisqip requirement introduces the largest effect on the observed decay-time distribution compared to other selection criteria.
In the second selection this track is combined with a second track to form a candidate for a \Dz decay into two hadrons (charge conjugate states are included unless stated otherwise).
The second track must have $\pt > 0.8\gevc$ and $\chisqip > 2$.
The decay vertex is required to have a flight distance \chisq per degree of freedom greater than $25$ and the \Dz invariant mass, assuming kaons or pions as final state particles, has to lie within $50\mevcc$ (or within $120\mevcc$ for a trigger whose rate is scaled down by a factor of $10$) around $1865\mevcc$.
The momentum vector of two-body system is required to point back to the $pp$ interaction region.

The event selection applies a set of criteria that are closely aligned to those applied at the trigger stage.
The final-state particles have to match particle identification criteria to separate kaons from pions~\cite{LHCb-DP-2012-003} according to their mass hypothesis and must not be identified as muons using combined information from the tracking and particle identification systems.

Flavour tagging is performed through the measurement of the charge of the pion in the decay \decay{\Dstarp}{\Dz\pip} (soft pion).
Additional criteria are applied to the track quality of the soft pion as well as to the vertex quality of the \Dstarp meson.
Using a fit constraining the soft pion to the $pp$ interaction vertex, the invariant mass difference of the \Dstarp and \Dz candidates, $\deltam$, is required to be less than $152\mevcc$.

About $10\,\%$ of the selected events have more than one candidate passing the selections, mostly due to one \Dz candidate being associated with several soft pions.
One candidate per event is selected at random to reduce the background from randomly associated soft pions.
The \Dz decay-time range is restricted to $0.25\ps$ to $10\ps$ such that there are sufficient amounts of data in all decay-time regions included in the fit to ensure its stability.

The whole dataset is split into four subsets, identified by the magnet polarity, and two separate data-taking periods to account for known differences in the detector alignment and calibration.
The smallest subset contains about $20\%$ of the total data sample.
Results of the four subsets are combined in a weighted average.

The selected events contain about $3.11\times10^6$ \decay{\Dz}{\Km\Kp} and $1.03\times10^6$ \decay{\Dz}{\pim\pip} signal candidates, where the \Dstarp meson is produced at the $pp$-interaction vertex, with purities of $93.6\%$ and $91.2\%$, respectively, as measured in a region of two standard deviations of the signal peaks in \Dz mass, $m(hh)$ (with $h=\PK,\Ppi$), and \deltam. 

\begin{figure}[bt]
\includegraphics[width=0.5\textwidth]{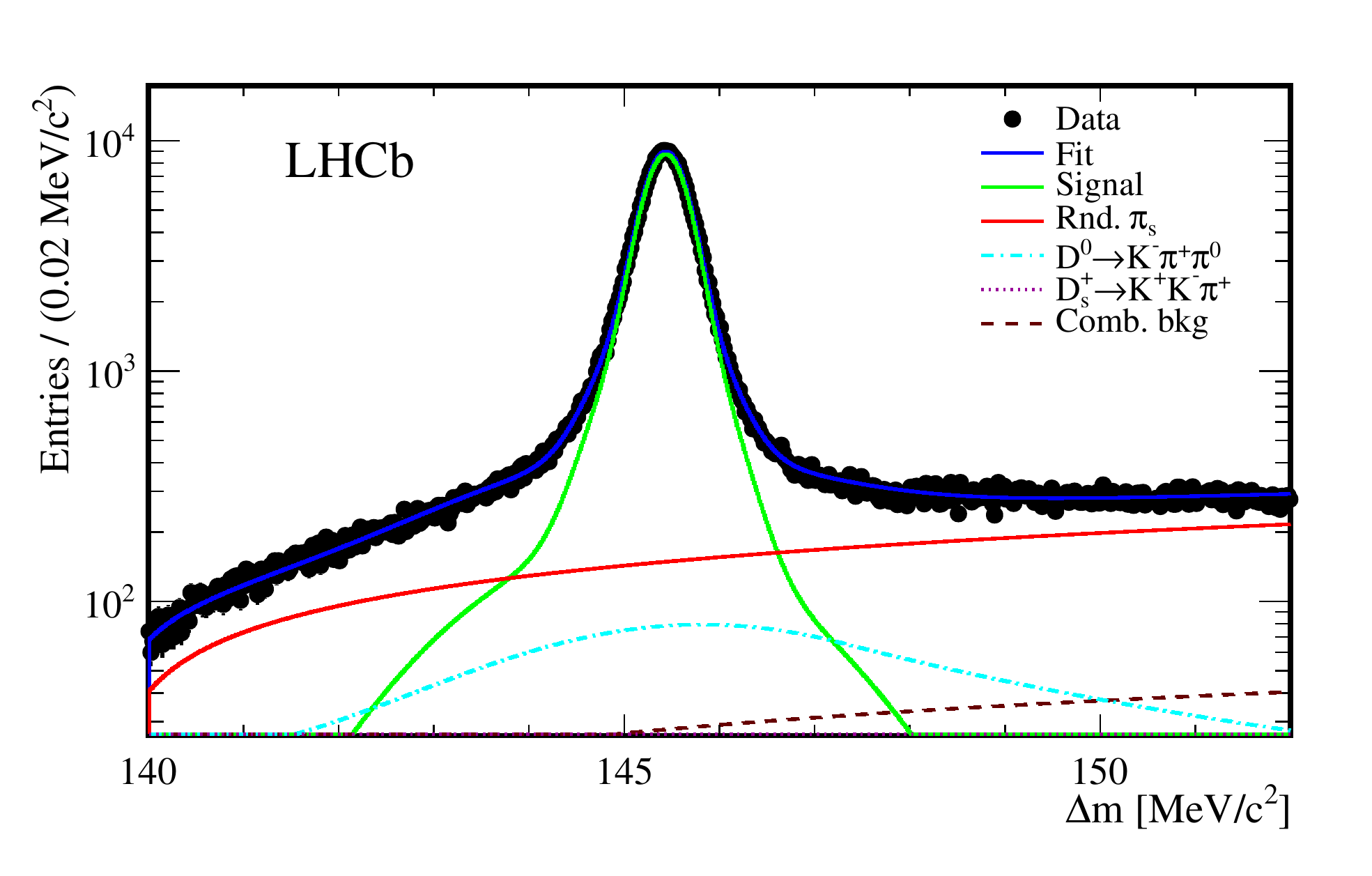}
\caption{\label{fig:mass}\small Fit of \deltam for one of the eight subsets, containing the \decay{\Dzb}{\Km\Kp} candidates with magnet polarity down for the earlier run period.}
\end{figure}

%%%%%%%%%%%%%%%%%%%%%%%%%%%%%%%%%%%%%%%%%%%%%%%%%%%%%%%%%%%%%%%%%%%%%%%%%%%%%%%%%%%%%%%%%

The effective lifetimes are extracted by eight independent multivariate unbinned maximum likelihood fits to the four subsamples, separated by the \Dz flavour as determined by the charge of the soft pion.
The fits are carried out in two stages, a fit to $m(hh)$ and \deltam to extract the signal yield and a fit to the decay time and $\ln(\chisqip)$ of the \Dz candidate to extract the effective lifetime.
The first stage is used to distinguish the following candidate classes: correctly tagged signal candidates, which peak in both variables; correctly reconstructed \Dz candidates associated with a random soft pion (labelled ``rnd. $\pi_{\mathrm{s}}$'' in figures), which peak in $m(hh)$ but follow a threshold function in \deltam; and combinatorial background.
The threshold functions are polynomials in $\sqrt{\deltam-m_{\pip}}$.
The signal peaks in $m(hh)$ and \deltam are described by the sum of three Gaussian functions.
For the $\pim\pip$ final state a power-law tail is added to the $m(hh)$ distribution to describe the radiative tail~\cite{Skwarnicki:1986xj}.
The combinatorial background is described by an exponential function in $m(hh)$ and a threshold function in \deltam. 

Partially reconstructed decays constitute additional background sources.
The channels that give significant contributions are the decays \decay{\Dz}{\Km\pip\piz}, with the charged pion reconstructed as a kaon and the \piz meson not reconstructed, and \decay{\Ds}{\Km\Kp\pip}, with the pion not reconstructed.
The former peaks broadly in \deltam while the latter follows a threshold function and both are described by an exponential in $m(hh)$.
Reflections due to incorrect mass assignment of the tracks are well separated in mass and are suppressed by particle identification and are not taken into account.
An example fit projection is shown in Fig.~\ref{fig:mass}.

Charm mesons originating from long-lived \bquark hadrons (secondary candidates) form a large background that cannot be separated in the mass fit.
They do not come from the interaction point leading to a biased decay-time measurement.
The flight distance of the \bquark hadrons causes the \Dz candidates into which they decay to have large \chisqip on average.
This is therefore used as a separating variable. 

Candidates for signal decays, where the \Dstarp is produced at the $pp$-interaction vertex, are modelled by an exponential function in decay time, whose decay constant determines the effective lifetime, and by a modified \chisq function in $\ln(\chisqip)$ of the form
\begin{equation}
\label{eq:lnchisq}
f(x) \equiv \begin{cases}
e^{ \alpha x - e^{ \alpha (x - \mu) } } & x \leq \mu \\
e^{ \alpha \mu + \beta (x - \mu) - e^{ \beta (x - \mu) } } & x > \mu, \\
\end{cases}
\end{equation}
where all parameters are allowed to have a linear variation with decay time.
The parameters $\alpha$ and $\beta$ describe the left and right width of the distribution, respectively, and $\mu$ is the peak position.
Secondary candidates are described by the convolution of two exponential probability density functions in decay time.
Since there can be several sources of secondary candidates, the sum of two such convolutions is used with one of the decay constants shared, apart from the smaller $\pim\pip$ dataset where a single convolution is sufficient to describe the data.
The $\ln(\chisqip)$ distribution of secondary decays is also given by Eq.~\ref{eq:lnchisq}, however, the three parameters are replaced by functions of decay time
\begin{equation}
\alpha(t)=A+B\, t+C\, \arctan(D\, t),
\end{equation}
and similarly for $\beta$ and $\mu$, where the parametrisations are motivated by studies on highly enriched samples of secondary decays and where $A$, $B$, $C$, and $D$ describe the decay-time dependence.

The background from correctly reconstructed \Dz mesons associated to a random soft pion share the same $\ln(\chisqip)$ shape as the signal.
Other combinatorial backgrounds and partially reconstructed decays for the $\Km\Kp$ final state are described by non-parametric distributions.
The shapes are obtained by applying an unfolding technique described in Ref.~\cite{Pivk:2004ty} to the result of the $m(hh)$, \deltam fit.
Gaussian kernel density estimators are applied to create smooth distributions~\cite{DesityEstimation}.

\begin{figure}[tb]
\includegraphics[width=0.5\textwidth]{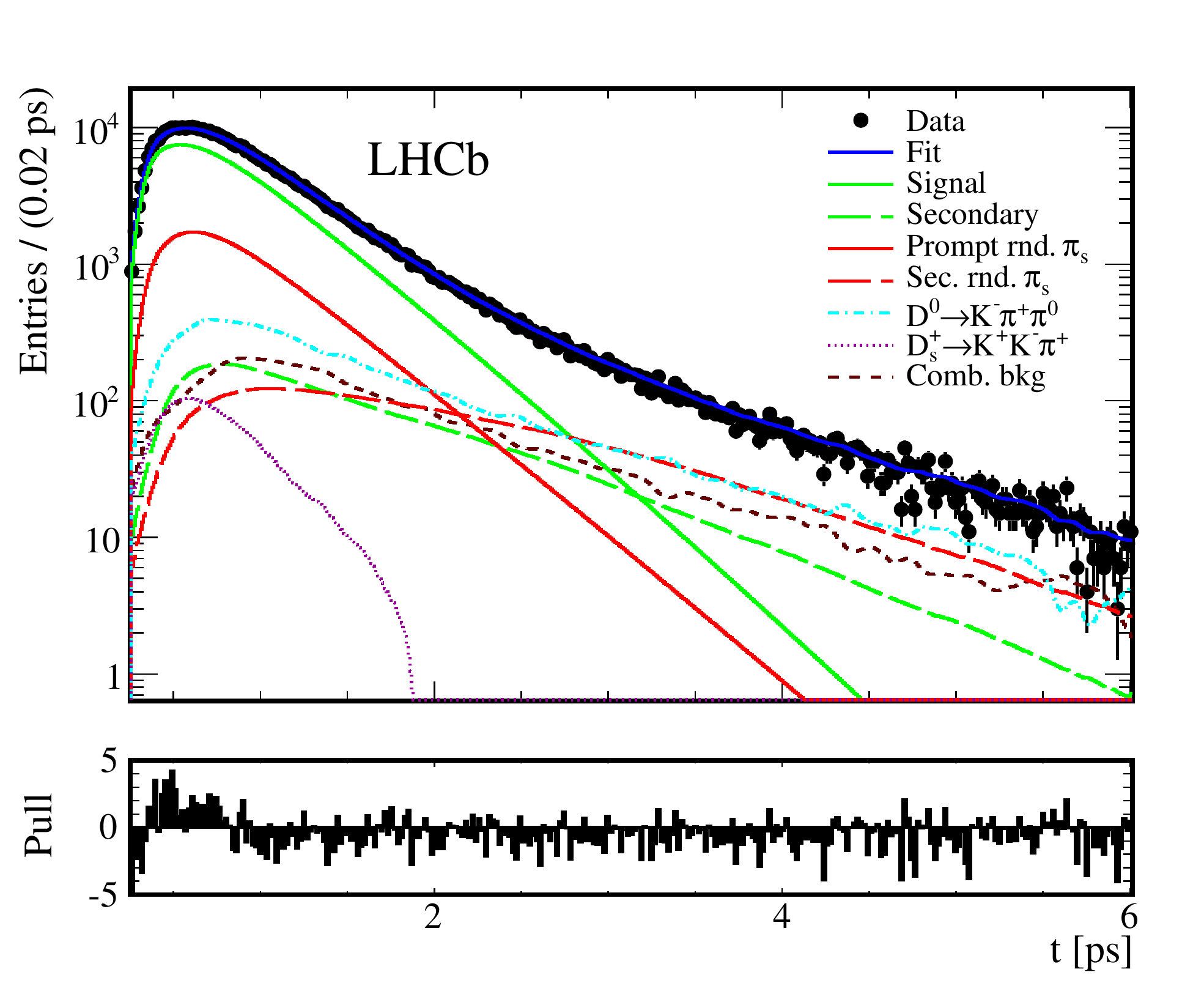}
\includegraphics[width=0.5\textwidth]{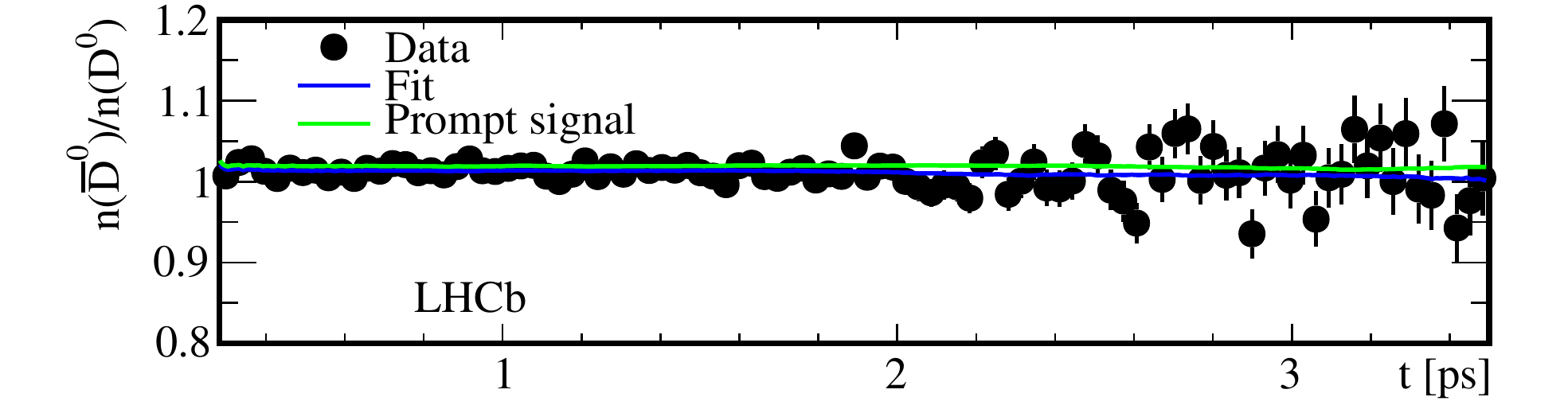}
\includegraphics[width=0.5\textwidth]{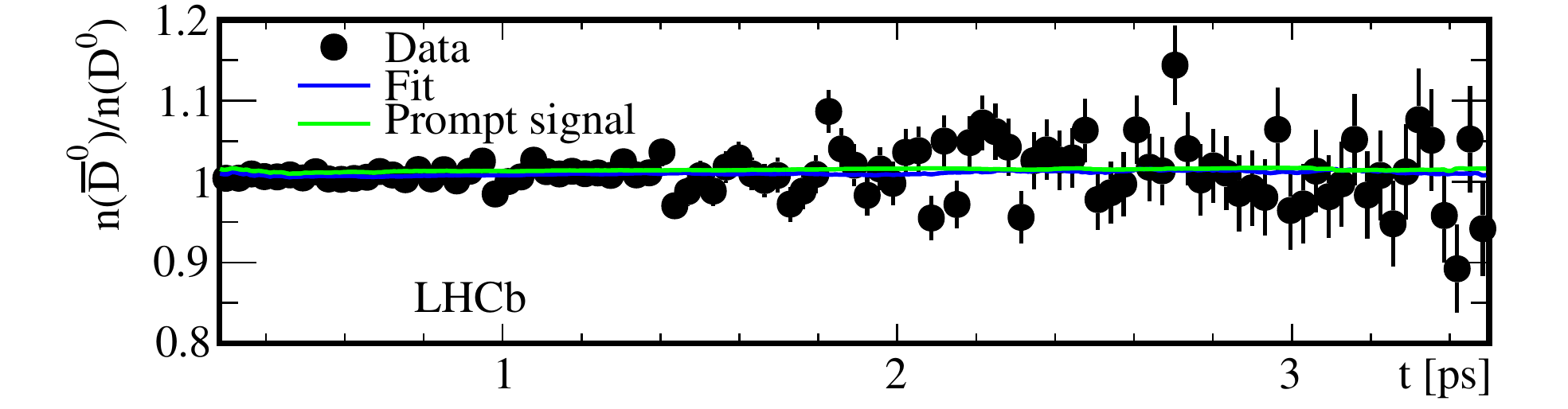}
\caption{\label{fig:time}\small (Top) Fit of decay time to \decay{\Dzb}{\Km\Kp} and corresponding pull plot for candidates with magnet polarity down for the earlier run period, where pull is defined as $({\rm data}-{\rm model})/{\rm uncertainty}$, and (middle and bottom) ratio of \Dzb to \Dz data and fit model for decays to $\Km\Kp$ and $\pim\pip$ for all data, respectively.}
\end{figure}

The detector resolution is accounted for by the convolution of a Gaussian function with the decay-time function. 
The Gaussian width is $50\fs$, an effective value extracted from studies of \decay{\PB}{\jpsi X} decays~\cite{LHCb-PAPER-2011-021}, which has negligible effect on the measurement.
Biases introduced by the selection criteria are accounted for through per-candidate acceptance functions which are determined in a data-driven way.
The acceptance functions, which take values of 1 for all decay-time intervals in which the candidate would have been accepted and 0 otherwise, enter the fit in the normalisation of the decay-time parametrisations.
The procedure for determination and application of these functions is described in detail in Refs.~\cite{LHCb-PAPER-2011-032,Gligorov:2012gr}.
Additional geometric detector acceptance effects are also included in the procedure.
An example decay-time fit projection is shown in Fig.~\ref{fig:time}.
The fit yields $\agamma(\PK\PK)=(-0.35\pm0.62)\times10^{-3}$ and $\agamma(\Ppi\Ppi)=(0.33\pm 1.06)\times10^{-3}$, with statistical uncertainties only.
The results of the four subsets are found to be in agreement with each other.

\begin{figure*}[tb]
\centering
\mbox{
\hspace{-0.02\textwidth}
\includegraphics[width=0.35\textwidth]{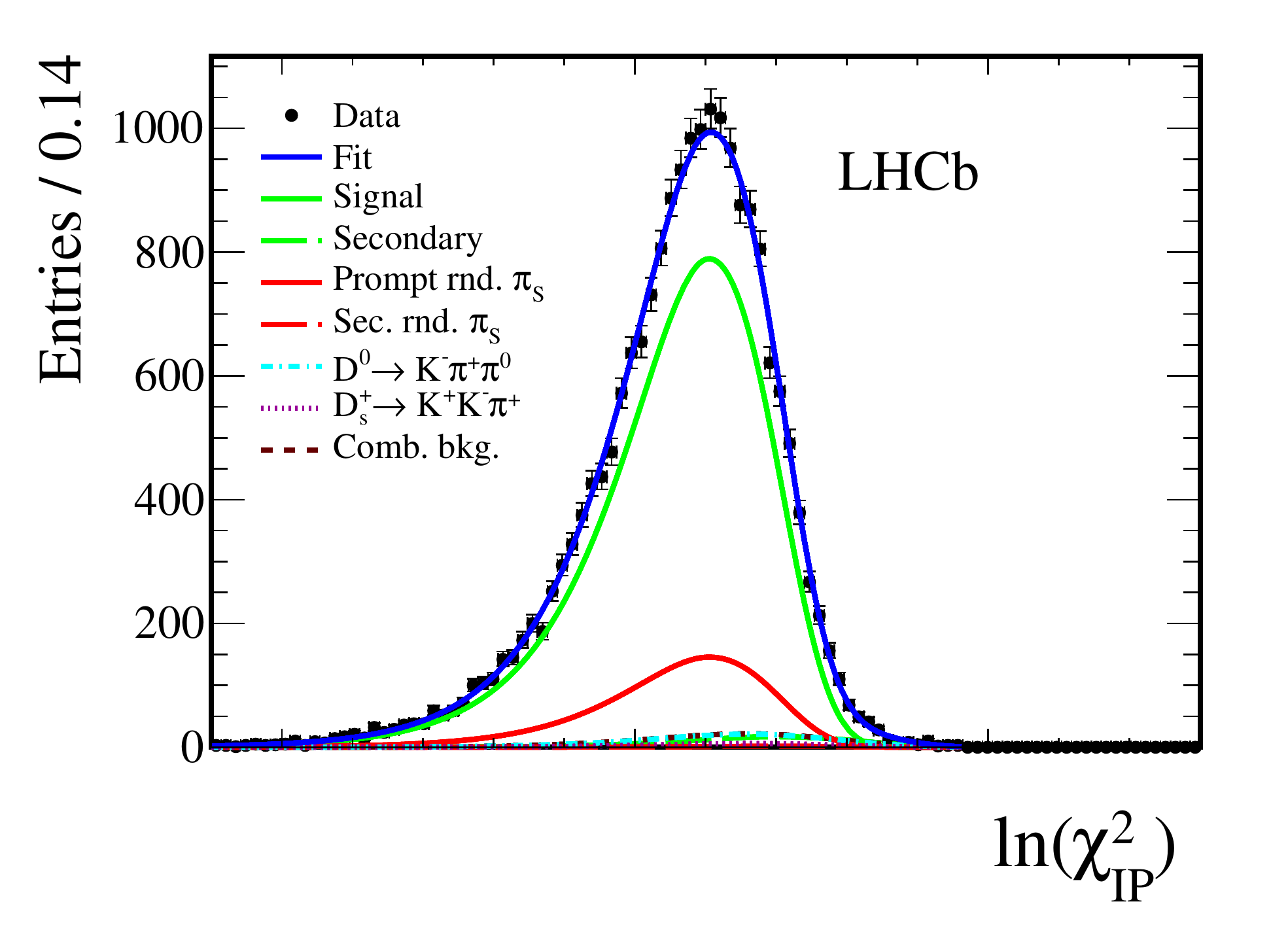}
\hspace{-0.03\textwidth}
\includegraphics[width=0.35\textwidth]{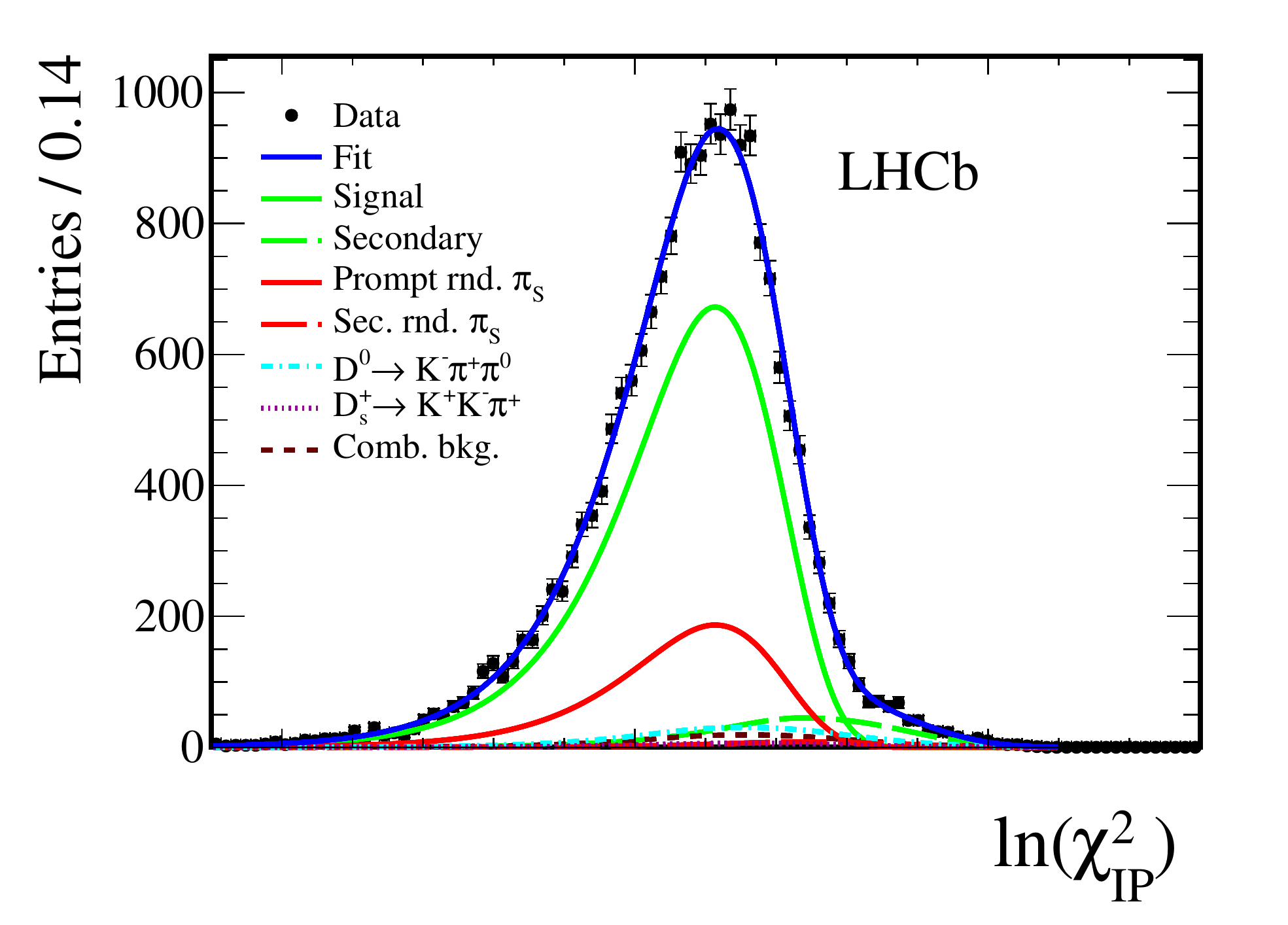}
\hspace{-0.03\textwidth}
\includegraphics[width=0.35\textwidth]{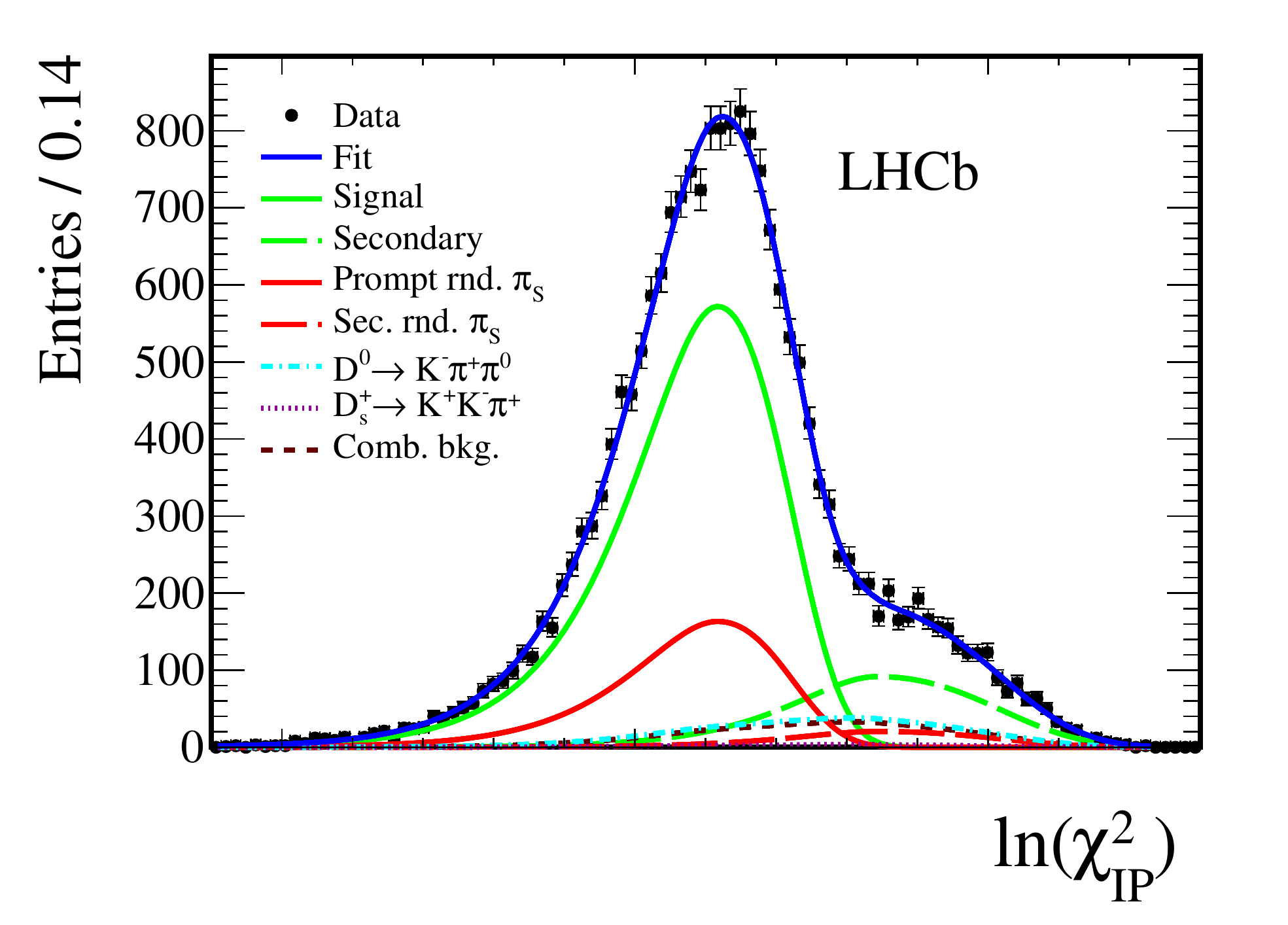}
}
\caption{\label{fig:binned_ipchi2}\small Fits of $\ln(\chisqip)$ for \decay{\Dzb}{\Km\Kp} candidates for decay-time bins (left to right) $0.25-0.37\ps$, $0.74-0.78\ps$, and $1.55-1.80\ps$.}
\end{figure*}

The fit has regions where the model fails to describe the data accurately, particularly at small decay times and intermediate values of $\ln(\chisqip)$ as shown in the pull plot in Fig.~\ref{fig:time}.
The same deviations are observed in pseudo-experiment studies, and are reproduced in several independent parametrisations, indicating that the origin is related to the non-parametric treatment of backgrounds in connection with non-ideal parametrisations of the $\ln(\chisqip)$ distributions. 
They do not significantly affect the central value of \agamma due to the low correlations between the effective lifetime and other fit parameters.
The deviations are very similar for fits to \Dz and \Dzb samples leading to their cancellations in the final asymmetry calculations as shown in Fig.~\ref{fig:time}.

In addition to the nominal procedure an alternative method is used, in which the data are binned in equally-populated regions of the decay-time distribution and the ratio of \Dzb to \Dz yields calculated in each bin.
This avoids the need to model the decay-time acceptance.
The time dependence of this ratio, $R$, allows the calculation of \agamma from a simple linear \chisq minimisation, with
\begin{equation}
R(t)\approx\frac{N_{\Dzb}}{N_{\Dz}}\left(1+\frac{2\agamma}{\tau_{\PK\PK}}t\right)\frac{1-e^{-\Delta t/\tau_{\Dzb}}}{1-e^{-\Delta t/\tau_{\Dz}}},
\end{equation}
where $\tau_{\PK\PK}=\tau_{\PK\Ppi}/(1+\ycp)$ is used as an external input based on current world averages~\cite{PDG2012,HFAG}, $N_{\Dzb}/N_{\Dz}$ is the signal yield ratio integrated over all decay times and $\Delta{}t$ is the bin width.
The dependence on $\tau_{\Dz}$ and $\tau_{\Dzb}$ cancels in the extraction of \agamma.
For this method the signal yields for decays, where the \Dstarp is produced at the $pp$-interaction vertex, for each decay-time bin are extracted by simultaneous unbinned maximum likelihood fits to $m(hh)$, \deltam, and $\ln(\chisqip)$.
Each bin is chosen to contain about $4\times10^4$ candidates, leading to $118$ and $40$ bins for $\Km\Kp$ and $\pim\pip$, respectively.
In general, the binned fit uses similar parametrisations to the unbinned fit, though a few simplifications are required to account for the smaller sample size per bin.
The evolution of the fit projections in $\ln(\chisqip)$ with decay time is shown in Fig.~\ref{fig:binned_ipchi2}.

The fits for both methods are verified by randomising the flavour tags and checking that the results for \agamma are in agreement with zero.
Similarly, the measurement techniques for \agamma are applied to the Cabibbo-favoured $\Km\pip$ final state for which they also yield results in agreement with zero.
The unbinned fit is further checked by comparing the extracted lifetime using the $\Km\pip$ final state to the world average \Dz lifetime, $(410.1\pm1.5)\fs$~\cite{PDG2012}.
The result of $(412.88\pm0.08)\fs$, where the uncertainty is only statistical, is found to be in reasonable agreement.
If the full difference to the world average were taken as a relative systematic bias it would lead to an absolute bias of less than $10^{-4}$ on \agamma.
Large numbers of pseudo-experiments, with both zero and non-zero input values for \agamma, are used to confirm the accuracy of the results and their uncertainties.
Finally, dependencies on \Dz kinematics and flight direction, the selection at the hardware trigger stage, and the track and vertex multiplicity, are found to be negligible.

%%%%%%%%%%%%%%%%%%%%%%%%%%%%%%%%%%%%%%%%%%%%%%%%%%%%%%%%%%%%%%%%%%%%%%%%%%%%%%%%%%%%%%%%%

The binned fit yields $\agamma(\PK\PK)=(0.50\pm0.65)\times10^{-3}$ and $\agamma(\Ppi\Ppi)=(0.85\pm 1.22)\times10^{-3}$.
Considering the statistical variation between the two methods and the uncorrelated systematic uncertainties the results from both methods yield consistent results.

%%%%%%%%%%%%%%%%%%%%%%%%%%%%%%%%%%%%%%%%%%%%%%%%%%%%%%%%%%%%%%%%%%%%%%%%%%%%%%%%%%%%%%%%%

The systematic uncertainties assessed are summarised in Table~\ref{tab:systematics}.
The effect of shortcomings in the description of the partially reconstructed background component in the $\Km\Kp$ final state is estimated by fixing the respective distributions to those obtained in fits to simulated data.
The imperfect knowledge of the length scale of the vertex detector as well as decay-time resolution effects are found to be negligible.
Potential inaccuracies in the description of combinatorial background and background from signal candidates originating from \bquark-hadron decays are assessed through pseudo-experiments with varied background levels and varied generated distributions while leaving the fit model unchanged.
The impact of imperfect treatment of background from \Dz candidates associated to random soft pions is evaluated by testing several fit configurations with fewer assumptions on the shape of this background.

\begin{table*}[thb]
\centering
\caption[Systematic uncertainties.]
{\small Systematic uncertainties, given as multiples of $10^{-3}$. The first column for each final state refers to the unbinned fit method and the second column to the binned fit method.}
\begin{tabular}{lcccc} 
\hline\hline
Source                                			      & $\agamma^{\rm unb}(KK)$ & $\agamma^{\rm bin}(KK)$ & $\agamma^{\rm unb}(\pi\pi)$ & $\agamma^{\rm bin}(\pi\pi)$\\
\hline
Partially reconstructed backgrounds 	                      & $\pm0.02$ & $\pm0.09$ & $\pm0.00$ & $\pm0.00$ \\
Charm from \bquark decays                                     & $\pm0.07$ & $\pm0.55$ & $\pm0.07$ & $\pm0.53$ \\
Other backgrounds                                             & $\pm0.02$ & $\pm0.40$ & $\pm0.04$ & $\pm0.57$ \\
Acceptance function                                           & $\pm0.09$ & ---       & $\pm0.11$ & ---       \\
Magnet polarity                                               & ---       & $\pm0.58$ & ---       & $\pm0.82$ \\
\hline
Total syst. uncertainty                                       & $\pm0.12$ & $\pm0.89$ & $\pm0.14$ & $\pm1.13$ \\
\hline\hline
\end{tabular}
\label{tab:systematics}
\end{table*}

The accuracy of the decay-time acceptance correction in the unbinned fit method is assessed by testing the sensitivity to artificial biases applied to the per-event acceptance functions.
The overall systematic uncertainties of the two final states for the unbinned method have a correlation of $0.31$.

A significant difference between results for the two magnet polarities is observed in the binned method.
As this cannot be guaranteed to cancel, a systematic uncertainty is assigned.
The unbinned method is not affected by this as it is not sensitive to the overall normalisation of the \Dz and \Dzb samples.
In general the two methods are subject to different sets of systematic effects due to the different ways in which they extract the results.
The systematic uncertainties for the binned method are larger due to the fact that the fits are performed independently in each decay-time bin.
This can lead to instabilities in the behaviour of particular fit components with time, an effect which is minimised in the unbinned fit.
The effects of such instabilities are determined by running simulated pseudo-experiments.

The use of the external input for $\tau_{\PK\PK}$ in the binned fit method does not yield a significant systematic uncertainty.
A potential bias in this method due to inaccurate parametrisations of other background is tested by replacing the probability density functions by different models and a corresponding systematic uncertainty is assigned.

%%%%%%%%%%%%%%%%%%%%%%%%%%%%%%%%%%%%%%%%%%%%%%%%%%%%%%%%%%%%%%%%%%%%%%%%%%%%%%%%%%%%%%%%%

In summary, the \CP-violating observable \agamma is measured using the decays of neutral charm mesons into $\Km\Kp$ and $\pim\pip$.
The results of 
$\agamma(KK)=(-0.35\pm0.62\pm0.12)\times 10^{-3}$
and
$\agamma(\pi\pi)=(0.33\pm1.06\pm0.14)\times 10^{-3}$,
where the first uncertainties are statistical and the second are systematic, represent the world's best measurements of these quantities.
The result for the $\Km\Kp$ final state is obtained based on an independent data set to the previous \lhcb measurement~\cite{LHCb-PAPER-2011-032}, with which it agrees well.
The results show no significant difference between the two final states and both results are in agreement with zero, thus indicating the absence of indirect \CP violation at this level of precision.

% Do not include this in analysis note and conference reports
\section*{Acknowledgements}

\noindent We express our gratitude to our colleagues in the CERN
accelerator departments for the excellent performance of the LHC. We
thank the technical and administrative staff at the LHCb
institutes. We acknowledge support from CERN and from the national
agencies: CAPES, CNPq, FAPERJ and FINEP (Brazil); NSFC (China);
CNRS/IN2P3 and Region Auvergne (France); BMBF, DFG, HGF and MPG
(Germany); SFI (Ireland); INFN (Italy); FOM and NWO (The Netherlands);
SCSR (Poland); MEN/IFA (Romania); MinES, Rosatom, RFBR and NRC
``Kurchatov Institute'' (Russia); MinECo, XuntaGal and GENCAT (Spain);
SNSF and SER (Switzerland); NAS Ukraine (Ukraine); STFC (United
Kingdom); NSF (USA). We also acknowledge the support received from the
ERC under FP7. The Tier1 computing centres are supported by IN2P3
(France), KIT and BMBF (Germany), INFN (Italy), NWO and SURF (The
Netherlands), PIC (Spain), GridPP (United Kingdom). We are thankful
for the computing resources put at our disposal by Yandex LLC
(Russia), as well as to the communities behind the multiple open
source software packages on which we depend.

\addcontentsline{toc}{section}{References}
\setboolean{inbibliography}{true}
\bibliographystyle{LHCb}
\bibliography{main,LHCb-PAPER,LHCb-CONF,LHCb-DP}
\end{document}